\documentclass[runningheads]{llncs}

\usepackage[T1]{fontenc}
\usepackage{amsmath}
\usepackage{amssymb}
\usepackage{booktabs}
\usepackage{url}
\usepackage[hidelinks]{hyperref}

\newcommand{\qlike}{\ensuremath{\mathrm{QLIKE}}}
\newcommand{\mselog}{\ensuremath{\mathrm{MSE}\text{-}\mathrm{log}}}
\newcommand{\VaR}{\ensuremath{\mathrm{VaR}}}
\newcommand{\E}{\ensuremath{\mathbb{E}}}

\newcommand{\rqTwoReduction}{75\%}
\newcommand{\rqTwoReductionCI}{[71\%, 78\%]}
\newcommand{\rqTwoReductionLow}{48\%}
\newcommand{\rqTwoReductionHigh}{86\%}

\newcommand{\coverageLow}{1.2\%}
\newcommand{\coverageHigh}{6.3\%}
\newcommand{\coverageCalLow}{3.5\%}
\newcommand{\coverageCalHigh}{3.6\%}

\newcommand{\coverageRemoved}{97\%}
\newcommand{\hmseCapital}{+69.5\%}
\newcommand{\proxySpearmanPark}{0.964}
\newcommand{\proxySpearmanRS}{1.000}
\newcommand{\kupiecSeries}{35}

\newcommand{\pairwiseRatioRaw}{2.91}
\newcommand{\pairwiseRatioRawCI}{[2.27, 3.25]}
\newcommand{\pairwiseRatioShared}{2.73}
\newcommand{\pairwiseRatioSharedCI}{[2.01, 3.44]}
\newcommand{\pairwiseRatioAligned}{0.67}
\newcommand{\pairwiseRatioAlignedCI}{[0.56, 0.99]}
\newcommand{\pairwiseRatioMSElog}{3.71}
\newcommand{\pairwiseRatioHMAE}{3.98}
\newcommand{\pairwiseRatioSharedAligned}{0.81}
\newcommand{\pairwiseRatioSharedAlignedCI}{[0.66, 1.22]}

\newcommand{\assetRatioRange}{1.44-4.55}

\newcommand{\bootstrapReplications}{1000}
\newcommand{\bootstrapBlockLength}{10}

\newcommand{\huberDelta}{1.0}

\renewcommand{\rqTwoReduction}{77\%}
\renewcommand{\rqTwoReductionCI}{[72\%, 79\%]}
\newcommand{\kupiecRejectRaw}{128}
\newcommand{\kupiecRejectAligned}{109}
\renewcommand{\kupiecSeries}{175}
\renewcommand{\rqTwoReductionLow}{59\%}
\renewcommand{\rqTwoReductionHigh}{88\%}
\renewcommand{\coverageLow}{1.3\%}
\renewcommand{\coverageHigh}{6.3\%}
\renewcommand{\coverageCalLow}{3.5\%}
\renewcommand{\coverageCalHigh}{3.6\%}
\renewcommand{\coverageRemoved}{97\%}
\renewcommand{\hmseCapital}{68.7\%}
\newcommand{\harRawDisp}{0.322}
\newcommand{\harAlignedDisp}{0.040}
\newcommand{\harReduction}{88\%}
\newcommand{\harLevelShare}{89\%}
\newcommand{\harxRawDisp}{0.307}
\newcommand{\harxAlignedDisp}{0.049}
\newcommand{\harxReduction}{84\%}
\newcommand{\harxLevelShare}{85\%}
\newcommand{\nharRawDisp}{0.246}
\newcommand{\nharAlignedDisp}{0.058}
\newcommand{\nharReduction}{76\%}
\newcommand{\nharLevelShare}{80\%}
\newcommand{\dlinearRawDisp}{0.186}
\newcommand{\dlinearAlignedDisp}{0.073}
\newcommand{\dlinearReduction}{61\%}
\newcommand{\dlinearLevelShare}{71\%}
\newcommand{\lgbmRawDisp}{0.121}
\newcommand{\lgbmAlignedDisp}{0.050}
\newcommand{\lgbmReduction}{59\%}
\newcommand{\lgbmLevelShare}{56\%}
\newcommand{\pinballAlignedBreach}{3.60\%}
\newcommand{\pinballRawBreach}{6.31\%}
\newcommand{\pinballMagnitude}{-21.3\%}
\newcommand{\huberAlignedBreach}{3.61\%}
\newcommand{\huberRawBreach}{6.23\%}
\newcommand{\huberMagnitude}{-20.6\%}
\newcommand{\mselogAlignedBreach}{3.57\%}
\newcommand{\mselogRawBreach}{5.92\%}
\newcommand{\mselogMagnitude}{-19.1\%}
\newcommand{\qlikeAlignedBreach}{3.59\%}
\newcommand{\qlikeRawBreach}{3.68\%}
\newcommand{\qlikeMagnitude}{+0.0\%}
\newcommand{\pattonAlignedBreach}{3.57\%}
\newcommand{\pattonRawBreach}{3.68\%}
\newcommand{\pattonMagnitude}{+0.8\%}
\newcommand{\hmaeAlignedBreach}{3.48\%}
\newcommand{\hmaeRawBreach}{2.19\%}
\newcommand{\hmaeMagnitude}{+24.7\%}
\newcommand{\hmseAlignedBreach}{3.55\%}
\newcommand{\hmseRawBreach}{1.28\%}
\newcommand{\hmseMagnitude}{+68.7\%}

\begin{document}

\title{Loss Choice or Model Choice? \\
The Role of Forecast Level in \\
Cryptocurrency Volatility Forecasting}
\titlerunning{Loss or Model Choice? Forecast Level in Crypto Volatility Forecasting}

\author{Andrzej Tokajuk\orcidID{0009-0006-7483-0770} \and
Jarosław A. Chudziak\orcidID{0000-0003-4534-8652}}
\authorrunning{A. Tokajuk, J. A. Chudziak}
\institute{Warsaw University of Technology, Warsaw, Poland \\
\email{\{andrzej.tokajuk.stud,jaroslaw.chudziak\}@pw.edu.pl}}

\maketitle

\begin{abstract}
Volatility forecasts play a central role in financial risk management because
their overall level and day-to-day movements affect downstream decisions.
Most studies compare forecasting models while keeping the training loss fixed.
Yet losses emphasize different errors and can target different properties of
future volatility, so raw comparisons may combine persistent forecast-level
differences with differences in daily forecast movements.
This leaves unresolved whether the importance of loss choice comes
mainly from the forecast level it targets or from differences that remain after
level adjustment. We address this gap through a comparison of
seven losses and five models across major cryptocurrencies. Validation-based
alignment adjusts the forecast level before the raw and aligned forecasts are
evaluated using statistical scores and one-day Value-at-Risk.
Before alignment, marginal score differences are larger across training losses. After alignment, they are larger across model configurations in the full comparison, while cross-loss differences in VaR breach rates narrow substantially.
Our contribution is a comprehensive evaluation of 
loss and model choice that
shows why losses can appear so influential in raw comparisons and how this
interpretation changes when forecast level and downstream risk are considered
explicitly.

\keywords{Loss functions \and Volatility forecasting
\and Cryptocurrency \and Value-at-Risk}
\end{abstract}

\section{Introduction}

Volatility forecasts translate market movements into estimates used for
position sizing, capital allocation and Value-at-Risk (VaR). Their usefulness
depends on their overall level and movements from one day to the
next. Forecasts can follow similar daily paths while implying substantially
different levels of financial risk.

Volatility forecasting research has mainly focused on model design, with the
training loss typically held fixed when competing models are evaluated. This
choice is not neutral: different losses can favour different quantities of the
conditional distribution \cite{gneiting2011}, leading the same model trained on
the same data to produce different forecasts. Raw score differences may thus
combine persistent forecast-level shifts with differences in daily forecast
movements. Because forecast level feeds directly into risk measures, this
ambiguity affects both statistical evaluation and practical use.

These considerations lead to three research questions:

\begin{enumerate}
\item[\textbf{RQ1}] How large is the effect of the loss function relative
to the effect of the forecasting model?
\item[\textbf{RQ2}] Does the loss mainly change the forecast level, or
does it also change its day-to-day dynamics?
\item[\textbf{RQ3}] How do these differences affect one-day VaR before and
after forecast-level alignment?
\end{enumerate}

We investigate these questions through a walk-forward comparison of seven
losses and five models across five major cryptocurrencies, holding non-loss
settings fixed. We compare marginal score variation across the two factors,
align forecast
levels using validation data, and evaluate raw and aligned forecasts using
statistical scores and one-day VaR.

The analysis shows that forecast level is central to interpreting differences
attributed to training loss, while the differences remaining after alignment
depend on the model. The VaR evaluation further shows that this distinction
carries into risk estimates. Our contribution is a comprehensive evaluation of
loss and model choice that clarifies what raw comparisons capture, what remains
after forecast-level adjustment, and why this distinction matters for
financial risk.

\section{Related Work}

Volatility forecasting research has primarily focused on model design, with
HAR remaining a strong baseline across horizons \cite{corsi2009}. Evidence is
mixed: machine-learning gains in U.S. equities are strongest at longer
horizons and with informative predictors \cite{christensen2023}. Broader
comparisons find no general advantage of nonlinear models over linear
alternatives \cite{branco2024}, while cryptocurrency results depend on the
asset, evaluation metric and forecast horizon \cite{dudek2024asc}.

In contrast, the role of the training loss has received much less attention
in volatility forecasting, although the broader forecasting literature shows
that it is not merely an optimisation choice. A loss determines which property
of the target distribution a model is trained to estimate, so forecasts
obtained with different losses need not represent the same conditional
quantity \cite{gneiting2011}. This distinction is particularly important for
volatility, which is not observed directly and must be approximated using a
proxy. Forecast rankings may therefore depend on both the evaluation score and
the volatility proxy \cite{hansen2006,patton2011}.

In related financial research, Transformer architectures have been compared
for stock selection \cite{kwiatkowski2026models}, while a separate study
examines how training losses affect portfolio outcomes within a fixed
Transformer \cite{kwiatkowski2025loss}. For volatility forecasting, QLIKE
training has improved finite-sample HAR forecasts \cite{puke2026coherent}.
These studies leave open how loss choice compares with model choice and how
much of the difference reflects forecast level. We examine both questions
and their implications for VaR.

\section{Methodology}

In this section, we define forecasts targeted by each loss, compare loss
and model effects, and use alignment to distinguish forecast level from
day-to-day dynamics.

\subsection{Loss Functions}

The training loss determines which errors a model avoids and hence which
feature of future variance it estimates. Let \(h>0\) denote the observed
variance proxy and \(f>0\) its one-day-ahead forecast. We use the log error
\(u=\log h-\log f\) and the variance ratio \(r=h/f\). An exact forecast gives
\(u=0\) and \(r=1\), while \(u>0\) or \(r>1\) indicates underprediction.

Each loss can favour a different forecast of \(h_{t+1}\) given
\(\mathcal F_t\) - we call this its forecast target. All means and medians in
Table~\ref{tab:objectives} are conditional on \(\mathcal F_t\). Huber-log uses
\(\rho_\delta(u)=u^2/2\) for \(|u|\leq\delta\) and
\(\delta(|u|-\delta/2)\) otherwise, with \(\delta=\huberDelta{}\). Here
\(m_\delta\) is the log-variance value that minimizes the conditional expected
Huber loss, and \(\exp(m_\delta)\) transforms it back to variance units.

\begin{table}[!ht]
\centering
\caption{Per-observation losses and the forecasts they target.}
\label{tab:objectives}
\scriptsize
\setlength{\tabcolsep}{3.1pt}
\begin{tabular}{llll}
\toprule
Group & Loss & \(L(h,f)\) & Forecast target \\
\midrule
Log error & \mselog & \(u^2\) & geometric mean \\
& Huber-log & \(\rho_\delta(u)\) & \(\exp(m_\delta)\) \\
& Pinball & \(|u|/2\) & median \\
\midrule
Mean target & \qlike & \(r-\log r-1\) & mean \\
& Patton-\(b\) & \(f-h+h\log r,\ b=-1\) & mean \\
\midrule
Relative error & HMAE & \(|r-1|\) & \(h\)-weighted median \\
& HMSE & \((r-1)^2\) & \(\E[h^2\mid\mathcal F_t]/\E[h\mid\mathcal F_t]\) \\
\bottomrule
\end{tabular}
\end{table}

MSE-log targets the conditional geometric mean, whereas Huber-log gives less
weight to large log errors. HMAE instead weights observations by \(h\). During
variance spikes, these losses can therefore favour different forecast levels.

\subsection{Comparison and Forecast-Level Alignment}

We compare all five models under all seven losses with fixed settings.
The primary comparison retains native predictors, so model choice includes
differences in inputs and history length; shared inputs provide a robustness
check.

Unequal numbers of losses and models make their ranges incomparable. We instead
average absolute differences over all 21 loss pairs and 10 model pairs. In block
$b$, let $q_{om}(b)$ be log mean test QLIKE for loss $o$ and model $m$.
Let $\bar q_o(b)$ average this score across models and $\bar q_m(b)$ across
losses. We define

\[
\Delta_L(b)=\frac{1}{21}\sum_{o<o'}
\left|\bar q_o(b)-\bar q_{o'}(b)\right|,\qquad
\Delta_M(b)=\frac{1}{10}\sum_{m<m'}
\left|\bar q_m(b)-\bar q_{m'}(b)\right|.
\]

The median of $\Delta_L(b)/\Delta_M(b)$ over 25 blocks summarizes marginal
score differences; values above one indicate greater differences across
losses. This descriptive comparison does not separate loss-model
interactions.

To isolate forecast level, we estimate one validation constant $c$ per
loss-model fit and apply it to test forecasts:

\begin{equation}
c=\frac{1}{|V|}\sum_{t\in V}\frac{h_t}{f_t},
\qquad
f^{\mathrm{aligned}}_t=c f_t,
\label{eq:calibration}
\end{equation}

where $V$ contains the validation dates. This $c$ minimizes validation QLIKE
over constant rescalings; using it on every date changes forecast level while
preserving relative movements.

For two losses within the same model and block, let
\(d_t=\log f_{o,t}-\log f_{o',t}\) with test average \(\bar d\). Subtracting
\(\bar d\) removes a constant multiplicative offset and leaves time-varying
differences. We define

\[
\mathrm{LevelShare}
=1-\frac{\sum_t(d_t-\bar d)^2}{\sum_t d_t^2}.
\]

Values near one indicate mainly level differences; values near zero indicate
time-varying differences. LevelShare uses test forecasts only and does not
determine the validation-based alignment constants.

\section{Experimental Setup}

This section presents the data, models, training and evaluation. Chronological
data and fixed non-loss settings help attribute differences to the training
loss rather than tuning.

\paragraph{Data and target.}
We use daily Binance USDT spot OHLCV data for BTC, ETH, BNB, XRP and ADA
through 31 May 2026. BTC and ETH begin on 5 October 2017, and the others at their listing dates.
Volatility proxies are computed before lagged predictors, so every input for
day $t+1$ is available by day $t$. The target is next-day unannualised
Garman-Klass variance \cite{garman1980}; Parkinson \cite{parkinson1980} and
Rogers-Satchell \cite{rogers1991} test measurement robustness.

\paragraph{Forecasting models.}
We compare HAR, HAR-X, NeuralHAR, DLinear and LightGBM. HAR \cite{corsi2009}
uses daily, weekly and monthly log variance; HAR-X adds returns, leverage,
alternative proxies, volume and lagged BTC variance for altcoins. NeuralHAR
and LightGBM \cite{ke2017lightgbm} share these predictors, whereas DLinear
\cite{zeng2023dlinear} uses 48-day target history. All models predict log
variance $\hat z$, transformed to $f=\exp(\hat z)$.

\paragraph{Training and walk-forward design.}
HAR and HAR-X are estimated deterministically; the other models use three seeds
and validation-based early stopping. Five expanding folds contain
1{,}013-2{,}529 training, 253 validation and 379 test days, with disjoint test
windows. Scalers use training data, validation controls early stopping and
alignment, and all 25 asset-fold blocks receive equal weight.

\paragraph{Evaluation and uncertainty.}
We use QLIKE from Table~\ref{tab:objectives} as the primary score; MSE-log,
HMAE and alternative proxies test robustness. Scores are averaged over test
dates and seeds within each asset-fold block. Confidence intervals use
\bootstrapReplications{} stationary bootstrap samples with expected block
length \bootstrapBlockLength{} and identical resampled dates across forecasts.

We use one-day 5\% Gaussian VaR, $\VaR_t=-q_{0.05}\sqrt{f_t}$, as a common
diagnostic of forecast-level effects, where $q_{0.05}$ is the standard normal
quantile. We compare raw and aligned magnitudes and breach rates; the Kupiec
test \cite{kupiec1995} assesses asset-level coverage.

\section{Results}

The results follow the three research questions. We first compare score
differences across losses and models, then determine how much of the loss
effect comes from forecast level rather than day-to-day dynamics, and finally
examine how these differences affect VaR.

\subsection{RQ1: Loss Choice Is a Major Source of Raw Variation}

To answer RQ1, the ratio from Section~3.2 equals one for equal marginal loss
and model variation. Its median is \pairwiseRatioRaw{} (95\% CI
\pairwiseRatioRawCI{}), making cross-loss differences about 2.9 times larger.
It remains above one with shared predictors
(\pairwiseRatioShared{}, 95\% CI \pairwiseRatioSharedCI{}) and across assets
(\assetRatioRange{}).

\begin{table}[!ht]
\centering
\caption{Mean test \qlike{} across 25 blocks; lower is better and bold marks
column minima. Stochastic models average three seeds.}
\label{tab:grid}
\scriptsize
\setlength{\tabcolsep}{3.4pt}
\begin{tabular}{lrrrrr}
\toprule
Loss & HAR & HAR-X & DLinear & N-HAR & LGBM \\
\midrule
\mselog
& 0.772
& 0.734
& 0.751
& 0.718
& 0.684 \\

Huber-log
& 0.812
& 0.771
& 0.786
& 0.746
& 0.703 \\

Pinball
& 0.836
& 0.785
& 0.808
& 0.763
& 0.713 \\

\qlike
& \textbf{0.592}
& \textbf{0.578}
& \textbf{0.594}
& \textbf{0.576}
& 0.637 \\

Patton-$b$
& 0.597
& 0.591
& 0.677
& 0.586
& \textbf{0.625} \\

HMAE
& 0.728
& 0.704
& 0.646
& 0.709
& 0.649 \\

HMSE
& 1.243
& 1.178
& 0.873
& 1.051
& 0.659 \\
\bottomrule
\end{tabular}
\end{table}

Table~\ref{tab:grid} reports every combination of loss and model. QLIKE and Patton generally lead
under QLIKE evaluation, whereas HMSE performs poorly for HAR and HAR-X but
remains competitive for LightGBM. Rankings are not universal because a score
favours its own target.

The loss-to-model ratio remains above one under MSE-log (\pairwiseRatioMSElog{}) and HMAE
(\pairwiseRatioHMAE{}). Relative to Garman-Klass, loss rankings have Spearman correlations
of \proxySpearmanRS{} with Rogers-Satchell and \proxySpearmanPark{} with Parkinson. Thus, the raw loss
effect is robust to both the evaluation score and the volatility proxy.

\subsection{RQ2: Forecast Level Explains Most Raw Loss Variation}

The ordering in Table~\ref{tab:grid} follows the targets in
Table~\ref{tab:objectives}: log-error losses produce the lowest levels, QLIKE
and Patton target conditional mean variance, and relative-error losses produce
higher levels. RQ2 tests whether these level differences drive the raw loss
effect.

\begin{table}[!ht]
\centering
\caption{Level share and cross-loss log-QLIKE variation by model; alignment
uses validation only.}
\label{tab:level}
\scriptsize
\setlength{\tabcolsep}{3.6pt}
\begin{tabular}{lrrrr}
\toprule
Model & Level share & Raw variation & Aligned & Reduction \\
\midrule
HAR
& \harLevelShare{}
& \harRawDisp{}
& \harAlignedDisp{}
& \harReduction{} \\

HAR-X
& \harxLevelShare{}
& \harxRawDisp{}
& \harxAlignedDisp{}
& \harxReduction{} \\

NeuralHAR
& \nharLevelShare{}
& \nharRawDisp{}
& \nharAlignedDisp{}
& \nharReduction{} \\

DLinear
& \dlinearLevelShare{}
& \dlinearRawDisp{}
& \dlinearAlignedDisp{}
& \dlinearReduction{} \\

LightGBM
& \lgbmLevelShare{}
& \lgbmRawDisp{}
& \lgbmAlignedDisp{}
& \lgbmReduction{} \\
\bottomrule
\end{tabular}
\end{table}

Level share is the fraction of squared log-forecast differences explained by a
constant offset. It is \harLevelShare{} for HAR but \lgbmLevelShare{} for
LightGBM, which retains more loss-specific daily variation.

Alignment reduces cross-loss variation by
\rqTwoReductionLow{}-\rqTwoReductionHigh{} across models and
\rqTwoReduction{} overall (95\% CI \rqTwoReductionCI{}).
The loss-to-model ratio decreases from
\pairwiseRatioRaw{} to \pairwiseRatioAligned{} (95\% CI
\pairwiseRatioAlignedCI{}), placing the aligned interval below one.

Ratios remain below one under MSE-log, HMAE and for every asset. With shared
predictors, the estimate is \pairwiseRatioSharedAligned{} (CI
\pairwiseRatioSharedAlignedCI{}), so the reversal is statistically clear only
in the full grid. This extends recent HAR evidence: QLIKE-training gains can
partly reflect movement toward the level favoured by the evaluation loss, not
better temporal learning alone \cite{puke2026coherent}.

\subsection{RQ3: Forecast-Level Differences Propagate to Value-at-Risk}

RQ3 moves from forecast comparisons to measured risk. Loss rankings under
QLIKE and MSE-log have Spearman correlation $-0.200$, so the preferred loss
depends on how quality is defined. Table~\ref{tab:var} tests whether the level
differences from RQ2 also change VaR magnitude and breach rates.

\begin{table}[!ht]
\centering
\caption{One-day 5\% Gaussian VaR by loss. Raw magnitude is relative to QLIKE;
bold marks breach rates closest to 5\%.}
\label{tab:var}
\scriptsize
\setlength{\tabcolsep}{3.2pt}
\begin{tabular}{lrrr}
\toprule
Loss & Raw magnitude & Raw breach rate & Aligned breach rate \\
\midrule
Pinball
& \pinballMagnitude{}
& \pinballRawBreach{}
& \pinballAlignedBreach{} \\

Huber-log
& \huberMagnitude{}
& \huberRawBreach{}
& \textbf{\huberAlignedBreach{}} \\

\mselog
& \mselogMagnitude{}
& \textbf{\mselogRawBreach{}}
& \mselogAlignedBreach{} \\

\qlike
& \qlikeMagnitude{}
& \qlikeRawBreach{}
& \qlikeAlignedBreach{} \\

Patton-$b$
& \pattonMagnitude{}
& \pattonRawBreach{}
& \pattonAlignedBreach{} \\

HMAE
& \hmaeMagnitude{}
& \hmaeRawBreach{}
& \hmaeAlignedBreach{} \\

HMSE
& \hmseMagnitude{}
& \hmseRawBreach{}
& \hmseAlignedBreach{} \\
\bottomrule
\end{tabular}
\end{table}

A calibrated 5\% VaR is breached on about 5\% of test days; lower rates are
more conservative. Before alignment, log-error losses produce smaller VaR and
more breaches, while HMSE is most conservative and \hmseCapital{} larger than
QLIKE. Rates span \coverageLow{}-\coverageHigh{}.

Alignment narrows this range to \coverageCalLow{}-\coverageCalHigh{} and
removes \coverageRemoved{} of its cross-loss variation. Most raw risk
differences thus arise from persistent forecast-level shifts rather than daily
signals, consistent with evidence that statistical and financial criteria need
not rank methods alike \cite{tokajuk2025pmformer}.

The Kupiec test rejects correct coverage in \kupiecRejectRaw{} of
\kupiecSeries{} asset-loss-model series before alignment and
\kupiecRejectAligned{} after it. Aggregate aligned breach rates remain below
5\%, so convergence across losses does not imply correct VaR calibration.
Alignment optimizes QLIKE rather than tail coverage; remaining discrepancies
may reflect the Gaussian assumption, the volatility proxy or forecast error.

\section{Discussion}

This study investigated whether cross-loss differences reflect
volatility dynamics or mainly forecast levels. A key finding is that
the balance depends on the model: forecast level explains most of the HAR
difference, whereas LightGBM shows more loss-specific daily variation. Raw
results capture the complete loss effect; aligned results show what remains
after validation-based level adjustment, including possible residual
test-period level differences.

The evidence covers five cryptocurrencies, one-day forecasts, five models and
seven losses, so the balance may differ across markets or horizons. Fixed
settings exclude loss-specific tuning, while constant alignment cannot correct
time-varying bias. VaR results also depend on OHLC-based proxies and a Gaussian
return model.
Raw and aligned results therefore answer complementary questions. Reporting
both indicates whether differences in QLIKE or VaR are primarily associated
with forecast level or day-to-day forecast differences.

\section{Conclusion and Future Work}

In this paper, we examined how the training loss shapes cryptocurrency
volatility forecasts relative to model choice. We found that much of the large
raw loss effect comes from forecast level rather than distinct dynamics,
although its importance depends on the model. 
After validation-based alignment,
marginal score differences are larger across model configurations in the
full grid, while the shared-input comparison remains inconclusive.
We also observed that alignment substantially narrows cross-loss differences in one-day VaR breach
rates, showing how the forecast level induced during training carries into a
downstream measure of financial risk.

The contribution of this work is a joint analysis of loss choice, model choice
and forecast level. By reporting raw and aligned results, the study explains
why a loss can appear superior under a statistical score without necessarily
learning better temporal dynamics. These findings offer a more transparent way
to interpret volatility benchmarks and their consequences for risk estimates.
Future work can extend the analysis to other markets and horizons,
loss-specific tuning and non-Gaussian risk models.

\end{document}